\documentclass[runningheads]{llncs}
\usepackage{graphicx}
\usepackage[hyphens]{url}
\usepackage{hyperref}
\usepackage[hyphenbreaks]{breakurl}
\usepackage[utf8]{inputenc}
\usepackage{orcidlink}
\hypersetup{
colorlinks=true, linkcolor=blue, filecolor=magenta, urlcolor=cyan,
}

\begin{document}
\title{Using Out-of-the-Box Frameworks for Contrastive Unpaired Image Translation for Vestibular Schwannoma and Cochlea Segmentation: An approach for the crossMoDA Challenge}
\titlerunning{Using Out-of-the-Box Frameworks for crossMoDA Challenge}
\author{Jae Won Choi\inst{1,2}\orcidlink{0000-0002-5937-7238}}
\authorrunning{JW Choi}
\institute{College of Medicine, Seoul National University, South Korea \and Department of Radiology, Armed Forces Yangju Hospital, South Korea \\ \email{djc0105@gmail.com}
}

\maketitle

\setcounter{footnote}{3}

\begin{abstract}
The purpose of this study is to apply and evaluate out-of-the-box deep learning frameworks for the crossMoDA challenge. We use the CUT model, a model for unpaired image-to-image translation based on patchwise contrastive learning and adversarial learning, for domain adaptation from contrast-enhanced T1 MR to high-resolution T2 MR. As data augmentation, we generate additional images with vestibular schwannomas with lower signal intensity. For the segmentation task, we use the nnU-Net framework. Our final submission achieved mean Dice scores of 0.8299 in the validation phase and 0.8253 in the test phase. Our method ranked 3rd in the crossMoDA challenge.
\end{abstract}

\section{Introduction}

Vestibular schwannomas (VS) are benign neoplasms of the nerve sheath most commonly occurring in the internal auditory canal and cerebellopontine angle. Currently, magnetic resonance (MR) imaging is the gold standard for diagnosis and surveillance of VS, and MR imaging protocols commonly include contrast-enhanced T1-weighted (ceT1) and high-resolution T2-weighted (hrT2) images \cite{lin2017management}. In general, smaller tumors are managed with long-term preservation rather than surgical resection \cite{goldbrunner2020eano}. Therefore, volumetric measurement of VS is critical, and a research group recently demonstrated a deep learning model for automatic segmentation of VS from ceT1 and hrT2 MR images \cite{shapey2019artificial}. However, the necessity of the MR contrast agent in the imaging of VS has been questioned, and an abbreviated MR imaging with only hrT2 images has been proposed as a cost-effective, faster, and safe alternative to the full MR with both ceT1 and hr T2 images \cite{lin2017management,buch2018noncontrast}. In this context, the crossMoDA (Cross-Modality Domain Adaptation for Medical Image Segmentation) challenge\footnote{\url{https://crossmoda.grand-challenge.org/}} has provided unpaired annotated ceT1 and non-annotated hrT2 images, a subset of a recently released publicly available imaging dataset \cite{shapey2021segmentation,https://doi.org/10.7937/tcia.9ytj-5q73}. The challenge participants were evaluated with results of segmentation of VS and the cochlea, a key anatomical structure in the treatment planning of VS, on hrT2 images.

Despite the recent rapid development of deep learning in medical imaging, few studies validate previous methods while most concentrate on novelty. The purpose of this study is to apply and evaluate out-of-the-box deep learning frameworks for the crossMoDA challenge. We use CUT\footnote{\url{https://github.com/taesungp/contrastive-unpaired-translation}} \cite{park2020contrastive}, a generic model for unpaired image-to-image translation based on patchwise contrastive learning and adversarial learning, to adapt ceT1 domain hrT2 domain. For the segmentation task in the hrT2 domain, we utilize nnU-Net\footnote{\url{https://github.com/MIC-DKFZ/nnUNet}} \cite{isensee2021nnu}, a framework that showed state-of-the-art performance in multiple medical image segmentation challenges \cite{isensee2019nnu}.

\section{Related Work}

Over the past few years, deep learning has been widely used for medical image segmentation, and many papers have shown great success in various tasks on different modalities \cite{isensee2019attempt,isensee2020nnunet,2021hecktor}. However, most of the recent high-performing deep learning models are based on supervised learning which often requires a large amount of carefully labeled data. The collection and annotation of data is especially challenging for medical image segmentation because of the high cost of pixel-level labeling by experts and heterogeneous nature of medical data. Thus, there has been many research efforts on learning with limited supervision although they have not been as successful as fully supervised learning \cite{peng2021medical}.
c
Domain adaptation (DA) is a popular subcategory of transfer learning that tackles limited supervision by utilizing labeled
data in source domains to execute tasks in a target domain. Unsupervised domain adaptation (UDA) refers to a domain adaptation task where only labeled data from the source domain and none from the target domain are available. Feature alignment is an approach for UDA that learns domain-invariant feature distributions across domains. One method of aligning feature spaces is through minimizing the discrepancy between the distributions based on measurements such as maximum mean discrepancy \cite{tzeng2014deep} and correlation alignment \cite{sun2016deep}. Also, the features spaces can be aligned via adversarial learning commonly based on domain classifier \cite{ganin2016domain,tzeng2017adversarial}.
Another line of research in UDA is the alignment of input spaces instead of features that makes use of unsupervised image-to-image translation. The popular strategy in this field is the cycle-consistency constraint of CycleGAN \cite{zhu2020unpaired} that inspired many networks such as UNIT \cite{liu2018unsupervised} and U-GAT-IT \cite{kim2020ugatit}, where bi-directional image translations are learned by two GANs. Some works address a one-sided image translation by utilizing some kind of content loss between domains. Benaim et al. \cite{benaim2017one} propose DistanceGAN that learns image translation by pairwise distances matching between images within domains. Fu et al. \cite{fu2019geometry} propose GCGAN to preserve the predefined geometric transformation between the input images before and after translation. Recently, Park et al. \cite{park2020contrastive} propose CUT to maximize the mutual information between the translations based on patchwise contrastive learning. Few studies have assessed application of CUT in medical image analysis. In this study, we use CUT to translate from ceT1 to hrT2 images.

\section{Methods}
Since we focus on applications of the publicly available frameworks, there is no modification to the mathematical setting or algorithm of the original works. All implementations were performed with PyTorch \cite{NEURIPS2019_9015} (version 1.7.1) on Nvidia RTX 3090 GPUs (single GPU training).

\subsection{Data}
The official training set includes ceT1 images with segmentation labels from 105 patients and hrT2 images without labels from a separate set of 105 patients. The VS (label 1) and cochlea (label 2) were manually segmented in consensus by the treating neurosurgeon and physicist using both the ceT1 and hrT2 images \cite{shapey2021segmentation}. As stated in the official challenge rules, no additional data was included for training.
The official validation set and test set include hrT2 MR images of 32 patients and 138 patients, respectively. The test phase is evaluated privately by the challenge organizers based on submissions using Docker containers during the evaluation phase of the challenge.

\subsection{Preprocessing}
Since the voxel spacings of the given training data are heterogeneous, we resample all cases to common voxel shaping of  0.6 × 0.6 × 1.0 mm. Labels were also interpolated likewise for the ceT1 domain. For each case, the input volume is scaled to [0.0, 1.0]. Then, a center $z-$axis is calculated as the average of $x$ and $y$ coordinates of voxels with intensity higher than the 75th percentile of the whole volume. We crop the input volume with a size of 256 × 256 pixels in $xy-$plane around the center $z-$axis, resulting in an image shape of 256 × 256 × $N$ voxels. Finally, we slice the volume data along the $z-$axis to acquire $N$ images with the size of 256 × 256 pixels because the CUT model only supports 2D images.

\begin{figure}[h]
\centering
   \includegraphics[width=12cm]{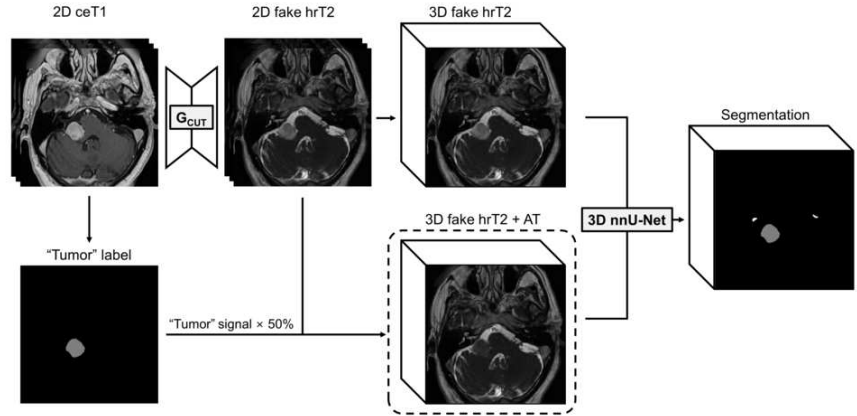}
   \hfil
\caption{Overview of our implementation of unpaired image translation with CUT and segmentation with nnU-Net. Training data is augmented by generating images with tumor signals reduced by 50\% (referred to as AT).}
\label{tab:figure-1}
\end{figure}

\subsection{Domain Adaptation}
We employ two model configurations in the official PyTorch implementation of CUT, CUT and FastCUT, to train models to perform DA from ceT1 to hrT2 domain on the training set. The training parameters for CUT and FastCUT were all based on the default options except that no resizing or cropping is performed and the number of epochs with the initial learning rate and the number of epochs with decaying learning rate are both set to 25. 

Although DA quality is eventually evaluated in the downstream segmentation task, objective visual quality assessment is also conducted using the widely used Fréchet Inception Distance (FID) metric, which measures the distance between the distributions of sets of images \cite{heusel2017gans}.

\subsection{Segmentation}
For the segmentation task, we use the default 3D full resolution U-Net configuration of the nnU-Net framework for training and inference except that the total epochs for training was set to 250.

We apply the trained DA model on all ceT1 images in the training set to acquire fake hrT2 images. The generated fake hrT2 images are stacked along the $z-$axis to reconstruct a volume data for each case in the training set. The fake hrT2 volumes and labels from the corresponding ceT1 images from the training set are used for training segmentation models. We hereafter refer to the nnU-Net model trained using fake hrT2 images generated by our trained CUT model as simply CUT, and likewise for FastCUT.

On MR T2 imaging, VS is generally hyperintense but some tumors can show heterogeneous signal intensity \cite{lin2017management}. To introduce heterogeneity of tumor signals to mimic such clinical characteristics, we generate additional training data by reducing the signal intensity of the labeled VS by 50\% (hereafter referred to as AT for "augmented tumor"). Thus, with AT, 210 cases were used as training data instead of 105 cases. We evaluate segmentation results of models trained on the original training data and the data with AT.

\section{Results}
\subsection{Domain Adaptation}
\begin{figure}[h]
\centering
   \includegraphics[width=12cm]{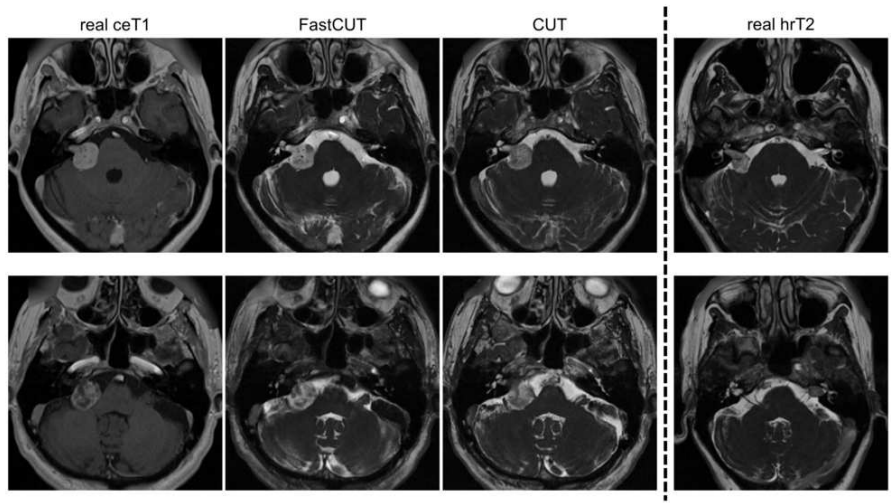}
   \hfil
\caption{Representative examples of UDA from ceT1 to hrT2. The second and third columns show fake hrT2 images generated by FastCUT and CUT from ceT1 images in the first column. For comparison, sample hrT2 images of different patients are presented in the last column.}
\label{tab:figure-2}
\end{figure}

The FID scores measured between real and fake hrT2 images were 32.85 for FastCUT and 11.15 for CUT. As shown in \autoref{tab:figure-2}, while both FastCUT and CUT achieved to translate from ceT1 images to hrT2 images, fine structures including the cochlea are better depicted by CUT.

\subsection{Segmentation}
All results for the segmentation task are obtained via the validation leaderboard of the crossMoDA challenge. Mean Dice scores are used to compare experiments, although other metrics including Dice scores and average symmetric surface distances (ASSD) for each label are also provided. All results are based on the ensembles of five-fold cross-validations on the training set. Ensembles are constructed by averaging softmax outputs.

\begin{figure}[h]
\centering
   \includegraphics[width=12cm]{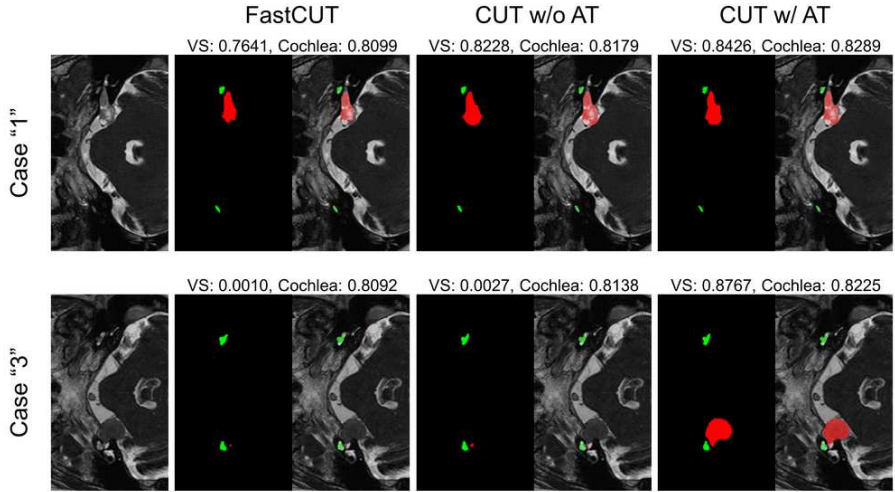}
   \hfil
\caption{Representative cases from validation set. Segmentation masks are displayed along with overlayed images and Dice scores for VS and cochlea. The first and second rows illustrate cases where tumors with cystic portions and darker signals, respectively, are better segmented by CUT with AT.}
\label{tab:figure-3}
\end{figure}

\autoref{tab:table-1} shows comparison of results on the validation set between segmentation models trained on images generated by FastCUT and CUT. Based on the higher performance of plain CUT over FastCUT, further evaluation of the effect of AT is conducted only on CUT. All metrics show better results for CUT with AT compared to plain CUT and FastCUT. Improvement of overall performance is mainly attributed to enhanced tumor segmentation as illustrated in \autoref{tab:figure-3}. The evaluation metrics showed improvements with AT not only for VS but also for cochlea even though AT involved only altering the signal intensity of the tumors.

\begin{table}[h]
\scriptsize
\centering\setlength{\tabcolsep}{2.5pt}
\renewcommand{\arraystretch}{1.2}
\caption{Comparison of results on the validation set between segmentation models trained on images generated by FastCUT, CUT without augmented tumor, and CUT with augmented tumor.}
\label{tab:table-1}
\begin{tabular}{l|c|c|c|c|c}
Experiment & Mean Dice              & Tumor Dice    & Tumor ASSD    & Cochlea Dice  & Cochlea ASSD  \\ \hline
FastCUT    & 0.7404±0.1514          & 0.6711±0.2941 & 5.5205±9.0409 & 0.8097±0.0256 & 0.1958±0.0399 \\
CUT w/o AT & 0.7703±0.1428          & 0.7217±0.2817 & 1.6655±1.8147 & 0.8188±0.0219 & 0.1765±0.0340 \\
CUT w/ AT  & \textbf{0.8299±0.0465} & 0.8375±0.0834 & 1.2940±1.2373 & 0.8223±0.0235 & 0.1720±0.0369
\end{tabular}
\end{table}

We submitted the ensemble of five-fold cross-validations of CUT with AT as the final submission for the challenge. Our method ranked 3rd in the test phase of the crossMoDA challenge with a mean Dice score of 0.8253. The mean Dice scores for tumor and cochlea were 0.8288 and 0.8217, respectively. The mean ASSD for tumor and cochlea were 1.0436 and 0.2858, respectively.

\section{Discussion}
In this work, we apply CUT, an unpaired image-to-image translation model, to generate fake hrT2 MR images from ceT1 MR images and nnU-Net, a framework for medical image segmentation, for segmentation of VS and cochlea on hrT2 MR images in the crossMoDA challenge. As data augmentation for the segmentation model, additional training data with lower tumor signals are generated. Our final submission achieved mean Dice scores of 0.8299 in the validation phase. In the test phase, with a mean Dice score of 0.8253, our method ranked 3rd in the crossMoDA challenge.

A significant strength of this study is that publicly available deep learning frameworks were applied on a public dataset without modifying default configurations as much as possible. Many publications on deep learning in medical image analysis focus on novel network architectures or training workflows to enhance performance. Also, they are often based on private datasets. However, specialized methods make it difficult for other researchers to reproduce the published results or apply them to different datasets or tasks. Thus, publicly available generic models may be a good choice of methods, especially in medical imaging where reproducibility and generalizability are critical to be actually used in clinical practice \cite{park2018methodologic}.

The current study has several limitations. First, it involves a limited number of experiments due to the circumstance of a challenge. Further experiments on different preprocessing and data augmentation may enhance performance. Also, comparison with other unsupervised image-to-image translation networks and out-of-the-box frameworks are warranted. Moreover, real hrT2 data were not used for training the segmentation model in this study. Self-training approach that retrains the segmentation model using pseudo-labels acquired from real hrT2 data would enhance segmentation results.

\section{Conclusion}
In conclusion, this study exploited a generic image-to-image translation network based on patchwise contrastive learning and adversarial learning to perform unsupervised domain adaptation for vestibular schwannoma and cochlea segmentation on high-resolution T2 MR images. Our results show that publicly available generic deep learning frameworks can achieve a certain degree of performance in medical imaging without a novel network or methodology. 
%
%
\bibliographystyle{splncs04}
\bibliography{myref}

\end{document}